# Evidence for Umklapp electron scattering emission from metal photocathodes

I-J. Shan, L.A. Angeloni, and W. Andreas Schroeder*

*Department of Physics, University of Illinois - Chicago*

**Abstract**

Comparison of the measured spectral emission properties of single-crystal Cu(001) and W(111) photocathodes to established photoemission theories reveal evidence for an additional one-photon emission process predominantly affecting electron emission near and below the photoemission threshold. This additional photoemission process is postulated to be due to a momentum-resonant Franck-Condon mechanism mediated by inelastic Umklapp electron scattering. An initial first-principles simulation of this emission process (involving the electron thermal effective mass, the inelastic electron mean free path at the vacuum level, and the number of Fermi surfaces in the metal), when combined with a direct one-step band emission model, is consistent with the measured spectral dependencies of both the quantum efficiency and mean transverse energy of electron photoemission from the two single-crystal metal photocathodes.

*Corresponding author: andreas@uic.edu

ORCIDs: I-J. Shan 0000-0003-1929-5593
L.A. Angeloni 0000-0003-2545-8168
W.A. Schroeder 0000-0002-1376-8401

## I. INTRODUCTION

The physics of photoemission from solid-state materials has been studied for over a century; from the first works of Hertz [1] and Lenard [2], their explanation by Einstein [3] and its subsequent verification by Millikan [4], to modern experimental methods such as angle-resolved photoelectron spectroscopy (ARPES) that map out energy-momentum relationship of bulk electronic states [5] and experiments verifying that the physics of photoemission occurs on a sub-femtosecond timescale [6]. Since 2009, the physics of photoemission from planar bulk metal photocathodes has been assumed to be well described by the Dowell-Schmerge (DS) formalism [7] modified by inclusion of a Fermi-Dirac electron distribution at finite temperature [8] and, if required, the effects of surface and chemical roughness on the spatial momentum distribution of emitted electrons [9,10]. More recently, a band-based direct one-step photoemission theory [11] incorporating the physics (dispersion and local density of states (DOS)) of both the emitting bulk and recipient vacuum states [12,13] has provided for better agreement with experiments on well-polished single-crystal metal photocathodes – removing the effects of both surface and chemical (i.e., work function variation) roughness [10]. While the former DS formalism does include the effects of electron scattering [14] on the emission quantum efficiency (QE), the role that *inelastic* electron-electron scattering plays in determining the QE and momentum distribution of emitted photoelectrons has not been treated in either analysis.

With the exception of a few experimental systems [15,16] the momentum distribution of a photoemitted electron beam is usually only measured in the transverse plane, providing the mean transverse energy (MTE) (or intrinsic emittance [7,17]) of the emitted electrons. For a cylindrically symmetric electron beam, the MTE is defined by $\langle p_T^2 \rangle/(2m_0)$, where $\Delta p_T = \sqrt{\langle p_T^2 \rangle}$ is the rms electron momentum transverse to the beam propagation direction (i.e., the normal to the photocathode emission face) and $m_0$ is the electron rest mass. Together with the QE, the MTE of electrons emitted from photocathodes is known to be of direct importance to the performance of x-ray free electron lasers (XFELs) [18-22], ultrafast electron diffraction (UED) systems [23-28], and dynamic transmission electron microscopes [29-31] – scientific instruments that are at the forefront of studies into the dynamic structural properties of solid-state materials [32-34] and molecules [35-38] with spatio-temporal resolution appropriate for atomic-scale motions. Substantial performance improvement of these instruments can generally only be achieved through enhanced beam brightness (proportional to the QE/MTE ratio [7,17]) and, most importantly, via MTE reduction as this leads to more efficient x-ray production in XFELs [39,40] and the resulting increase in transverse beam coherence improves the fidelity of electron diffraction patterns [41,42]. Consequently, a fundamental understanding of the electron emission processes governing the physics of photoemission is a key step to guide the selected discovery (or fabrication) of the desired future high brightness and low intrinsic emittance photocathodes [17,43-45].

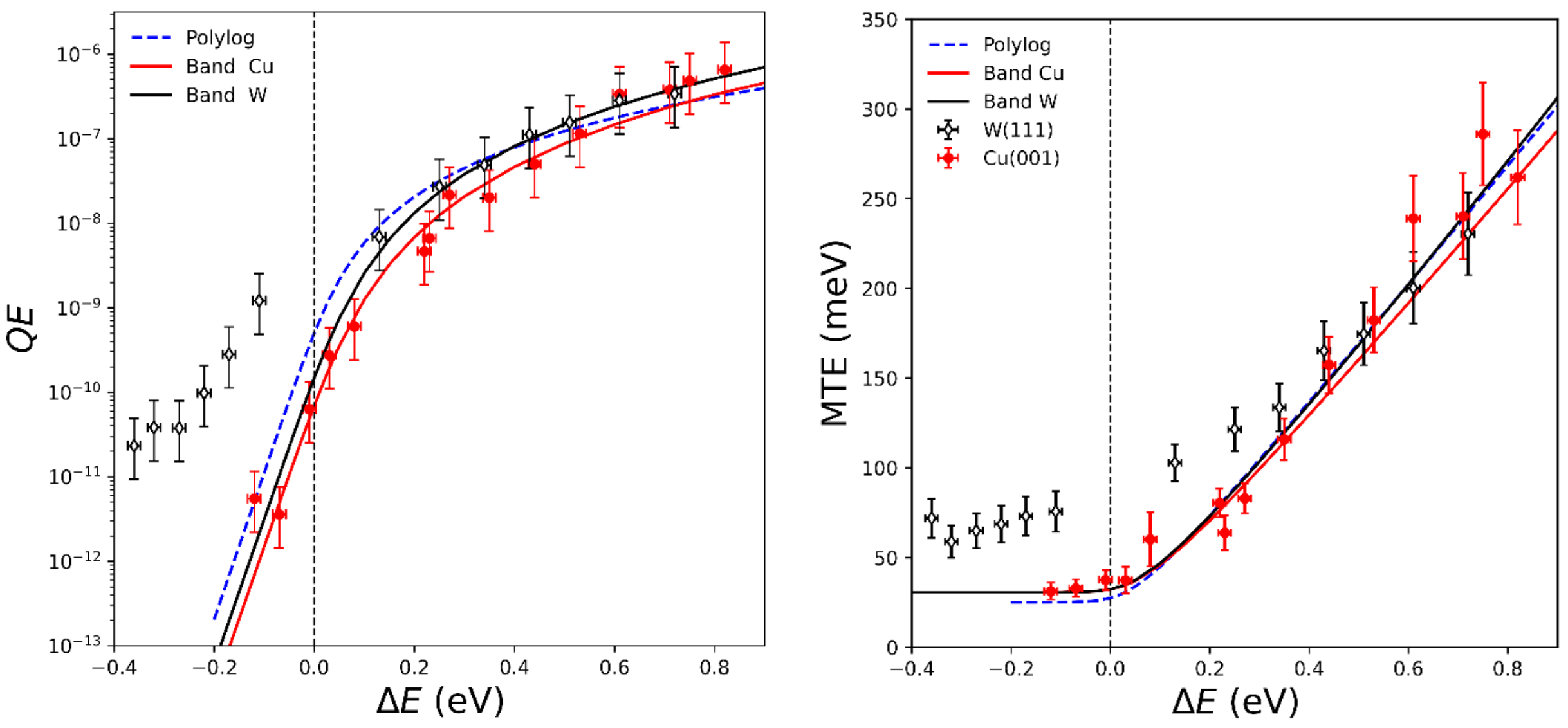

(a) (b)

FIG 1. The measured (a) quantum efficiency (QE) and (b) mean transverse energy (MTE) of electron emission as a function of excess photoemission energy $\Delta E$ for single-crystal Cu(001) (red data points) and W(111) (black data points) photocathodes. The predicted spectral trends from a one-step direct band emission simulation are shown as the solid red and black line respectively. Also plotted as a blue dashed line are the predicted trends for the QE and MTE from the temperature-extended DS formalism [8] (equations (1) and (2)).

A series of experimental investigations on polished and high-purity single-crystal metal photocathodes all with sub-10 nm rms surface roughness has revealed that their near and sub-threshold, single-photon, photo-emissive properties do not conform to either the temperature-extended Dowell-Schmerge (DS) formalism [7,8] or the more recent one-step direct band emission model [12,13]. An example of the observed discrepancy between experimental measurements and current theoretical expectations is shown in Figure 1. The figure displays the quantum efficiency (QE) and mean transverse energy (MTE) of the emitted electrons for two representative photocathodes (Cu(001) and W(111)) as a function of the excess photoemission energy, $\Delta E = \hbar\omega - \phi$, where $\hbar\omega$ is the incident photon energy and $\phi$ is the surface work function. Of immediate note is that the temperature-extended DS model [7,8] at 300 K (dashed lines) is not consistent with the experimental data: It forecasts an MTE of 25 meV below threshold ($\Delta E < 0$) (Figure 1(b)) which, although close to the ~29 meV measured for Cu(001), is certainly in disagreement with the W(111) measurements, and its predicted overall spectral QE trend is not completely in agreement with either set of measurements (Figure 1(a)). This is primarily because the DS analysis does not include the physics (i.e., dispersion and DOS) of either the emitting bulk electronic states in the metal photocathodes or the recipient vacuum states; for example, the square root of the energy ($\sqrt{E}$) dependence of the vacuum DOS biases the MTE to higher than expected values. On the other hand, the one-step direct band-based emission model (solid lines) [12], which includes the vacuum DOS, provides for a reasonable interpretation of the spectral dependence of both the QE and MTE for the Cu(001) photocathode; but it fails below threshold for the W(111) photocathode.

In particular, the measured below threshold QE of the W(111) photocathode strongly deviates from theoretical expectations, as evidenced by the 4-5 order of magnitude discrepancy at $\Delta E \approx -0.2$ eV (Figure 1(a)). Further, the measured ~70 meV below threshold MTE for W(111) is a factor of more than two greater than the ~30 meV predicted for direct one-step emission from the 300 K Boltzmann tail populating the emitting band in the relevant Γ-P direction of the Brillouin zone (Figure 1(b)).

The observed inconsistencies between current photoemission models and the experimental data point to an omission, in our theoretical understanding of photoemission from metals, of an electron emission mechanism that is strongest, relative to the direct band emission [12], predominantly near and below the photoemission threshold. Moreover, the strength of such an additional emission process is evidently material dependent, potentially related to differences in the electronic properties of metals. In this paper, we postulate that this is indeed the case and propose that the additional emission is due to momentum-resonant Franck-Condon (FC) emission [44,46-48] mediated by inelastic Umklapp electron scattering (UES) [49]. After a description of the experimental methods used to obtain the data displayed in Figure 1, we discuss the incorporation of the basic FC-UES emission physics into our band-based photoemission model [12,13]. The resulting combined theory is then compared to the QE and MTE measurements for the Cu(001) and W(111) photocathodes. Thereafter, the limitations of our simplified theoretical approach are discussed with a view to providing a pathway for a more sophisticated simulation of the observed effect.

## II. RESEARCH METHODS

### A. Experimental details

The detailed 300 K spectral measurements of the QE and MTE presented in Figure 1 were obtained using a photocathode characterization system that has been described fully elsewhere [13,50]. Briefly, the experimental system consists of a vacuum system (~$10^{-8}$ torr) housing a 10-20 kV DC electron gun (with a 1 cm acceleration gap) into which the photocathode samples are mounted with their surface normal parallel to the gun axis. The photocathode is illuminated by tunable ($\hbar\omega \approx 3$ to 5.3 eV) sub-picosecond ultraviolet (UV) radiation produced by the sum frequency up-conversion of the signal and idler radiation generated by the optical parametric amplification (OPA) of a continuum [13,50], all driven by a front-end 20 W, 16.7 MHz, femtosecond Yb:fiber laser system (aeroPULSE FS20, NKT Photonics). The *p*-polarized, 230-400 nm radiation with an average power of 10-100 μW is incident at 60° and focused to an irradiation area of $2\pi\sigma_x\sigma_y \approx 10^{-4}$ cm$^2$ (a near-Gaussian elliptical beam with rms widths of $\sigma_x \cong 50$ μm and $\sigma_y \cong 25$ μm) on the metal photocathode surface. The consequent incident absorbed pulse fluence of less than 100nJ/cm$^2$ over the typical ~10nm UV absorption depth in good metals with an electron density greater than $10^{22}$ cm$^{-3}$ results in less than a 1 K increase in the temperature of the photocathode's electron distribution.

After exiting the DC gun, the accelerated photoelectrons enter a 42 cm drift region at the end of which they are detected using a 10 μm pore dual micro-channel plate (MCP) coupled to a P-43 phosphor screen (BOS-18, Beam Imaging Solutions). The optical output (phosphorescence) is imaged with 8:5 demagnification onto a 2.4 μm pixel CMOS digital camera (FL-20BW, Axoim Optics). Detailed particle tracking through the experimental vacuum system and subsequent optical imaging reveals a 0.00214 $(m_0.eV)^{1/2}$/pixel transverse momentum calibration at the employed 16 kV gun voltage, which agrees with an Analytical Gaussian (AG) beam propagation simulation [51,52] of the photocathode characterization system. For the 25 μm point spread function (PSF) of the phosphor and the estimated 5 μm resolution of the achromat-based *f*/2 optical imaging system, this gives the experimental system a theoretical MTE resolution limit of less than 0.5 meV [50]. Together with the 25 μm PSF of the P-43 phosphor, the 42 cm drift region ensures that the ellipticity of the incident laser beam (i.e., the emission source) does not significantly affect the extraction of the MTE from the measured far-field electron beam profiles; for example, for a nominally thermally-limited MTE of 25 meV at 300 K, the ellipticity of the detected beam is less than 2% – smaller than the typical ±10% uncertainty in the measurements. The QE of photoemission is determined through a calibration of the signal amplification voltages on the MCP/phosphor detector using the known photoemission efficiency of a reference Rh(110) photocathode [13] (calibrated using a Faraday cup) and a measurement of the incident UV laser power.

The 99.999% purity single-crystal metal photocathodes were obtained from Princeton Scientific [53], are oriented to within 1° of the specified crystallographic direction and were polished to a rms surface roughness of less than 10 nm. The polish is sufficient to ensure that surface roughness effects do not affect the MTE measurements [9,10]. Prior to insertion in the DC photoelectron gun, the metal photocathodes are cleaned using propan-2-ol and n-hexane to remove physical surface contaminants and oils. Preceding this cleaning, the Cu(001) photocathode is exposed to 5% acetic acid solution for 15-20 minutes to remove most of its native surface oxide layer [54]. After insertion into the DC gun, both the studied Cu(001) and W(111) photocathodes are subjected to additional 'laser cleaning' at the 257 nm ($\hbar\omega$ = 4.81 eV), the fourth harmonic of the Yb:fiber laser, for ~30 minutes at a laser power of around 100 μW and a 4× larger ~50×100 μm rms incident spot size on the photocathode surface. This laser cleaning process [50,55-57] with UV-C radiation is terminated once the associated increase in the MTE is observed to cease, indicating that the surface work function has stabilized [7]. We note that the employed ~1W/cm$^2$ average incident laser irradiance and the equivalent low ~60nJ/cm$^2$ incident laser pulse fluence for our 16.7MHz repetition rate laser system are both insufficient to cause ablation of (or damage to) the metal photocathode surfaces [56].

### B. Prior photoemission models

The two theoretical photoemission formalisms displayed by the solid and dashed lines in Figure 1 have also been described in detail elsewhere [7,8,12,13]. The temperature-extended DS formalism (plotted as the dashed lines) is an emission model that incorporates the Fermi-Dirac distribution describing state occupation at finite temperature into the 'zero temperature' Dowell-Schmerge photoemission analysis [7]. As a result, it does not include the DOS of either the emitting bulk photocathode or recipient vacuum states. As shown by Vecchione et al. [8], this results in universal and straightforward PolyLog-based expressions for both the QE and MTE that are independent of the electronic properties of the photocathode material;

$$QE \propto \frac{Li_2\left[-exp\left(\frac{\Delta E}{k_B T_e}\right)\right]}{Li_2\left[-exp\left(\frac{\varepsilon_F}{k_B T_e}\right)\right]} \tag{1}$$

$$MTE = \frac{Li_3\left[-exp\left(\frac{\Delta E}{k_B T_e}\right)\right]}{Li_2\left[-exp\left(\frac{\Delta E}{k_B T_e}\right)\right]} . k_B T_e \tag{2}$$

where $Li_n$ are PolyLog functions of order $n$, $\varepsilon_F$ is the Fermi energy, $k_B$ is Boltzmann's constant, and $T_e$ is the electron temperature which is placed equal to the 300 K lattice temperature for the dashed lines in Figure 1. Notable is that as the temperature approaches zero, these expressions recover the $\Delta E$ dependencies of the Dowell-Schmerge photoemission formalism [7]; namely, $QE \propto \Delta E^2$ and $MTE = \Delta E/3$. Further, below threshold, the MTE approaches $k_B T_e$ since it is only the Boltzmann tail of the Fermi-Dirac distribution that is above the vacuum level and hence can emit.

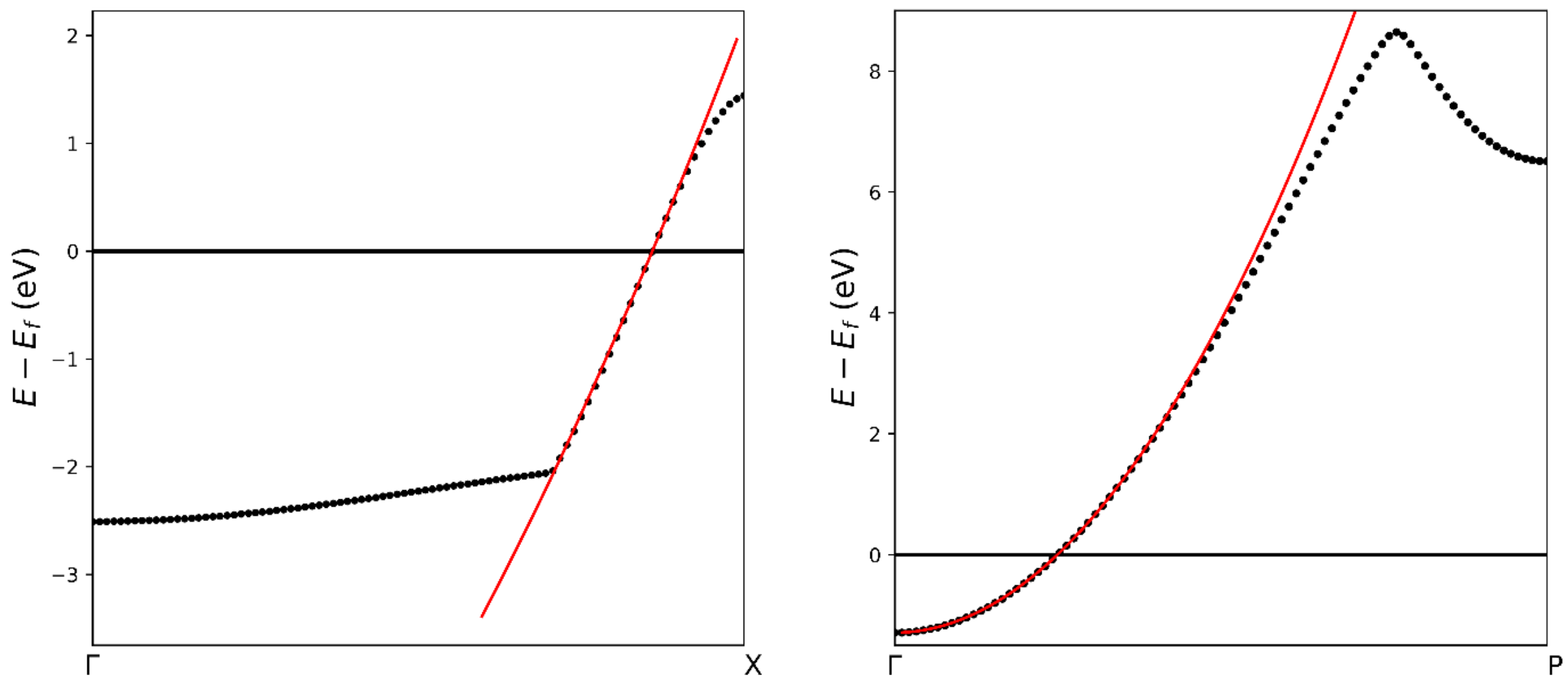

(a) (b)

FIG 2. Parabolic longitudinal band fits (red lines) to the DFT evaluated dispersion (black dots) of the emitting band for (a) Cu(001) in the Γ-X direction and (b) W(111) in the Γ-P direction.

In contrast, the one-step direct band photoemission model [12,13] incorporates the physics of both the emitting electronic band states in the solid-state photocathode and the recipient vacuum states. Briefly, in common with the DS formalism, the transverse momentum, $p_T$, of the emitted electron is conserved [5] but its value (at any energy) is determined by the dispersion characteristics of the emitting band state(s). This means that the employed one-step direct photoemission formalism can only be compared to single-crystal photocathodes whose electronic band structure is known. For the Cu(001) and W(111) photocathodes discussed in this paper, the electronic band structure around the Fermi level in the crystallographic emission direction is evaluated by density functional theory (DFT) and the result compared to the literature to assure accuracy; for copper the work of Burdick [58] and Marini et al.[59], and for tungsten the work of Christensen and Feuerbacher [60]. Our *Ab initio* band structure calculations employ the Amsterdam Modeling Suite's BAND module [61] with functionals from the LibXC library [62]; Cu using rSCAN [63] and W using PBEsol [64].

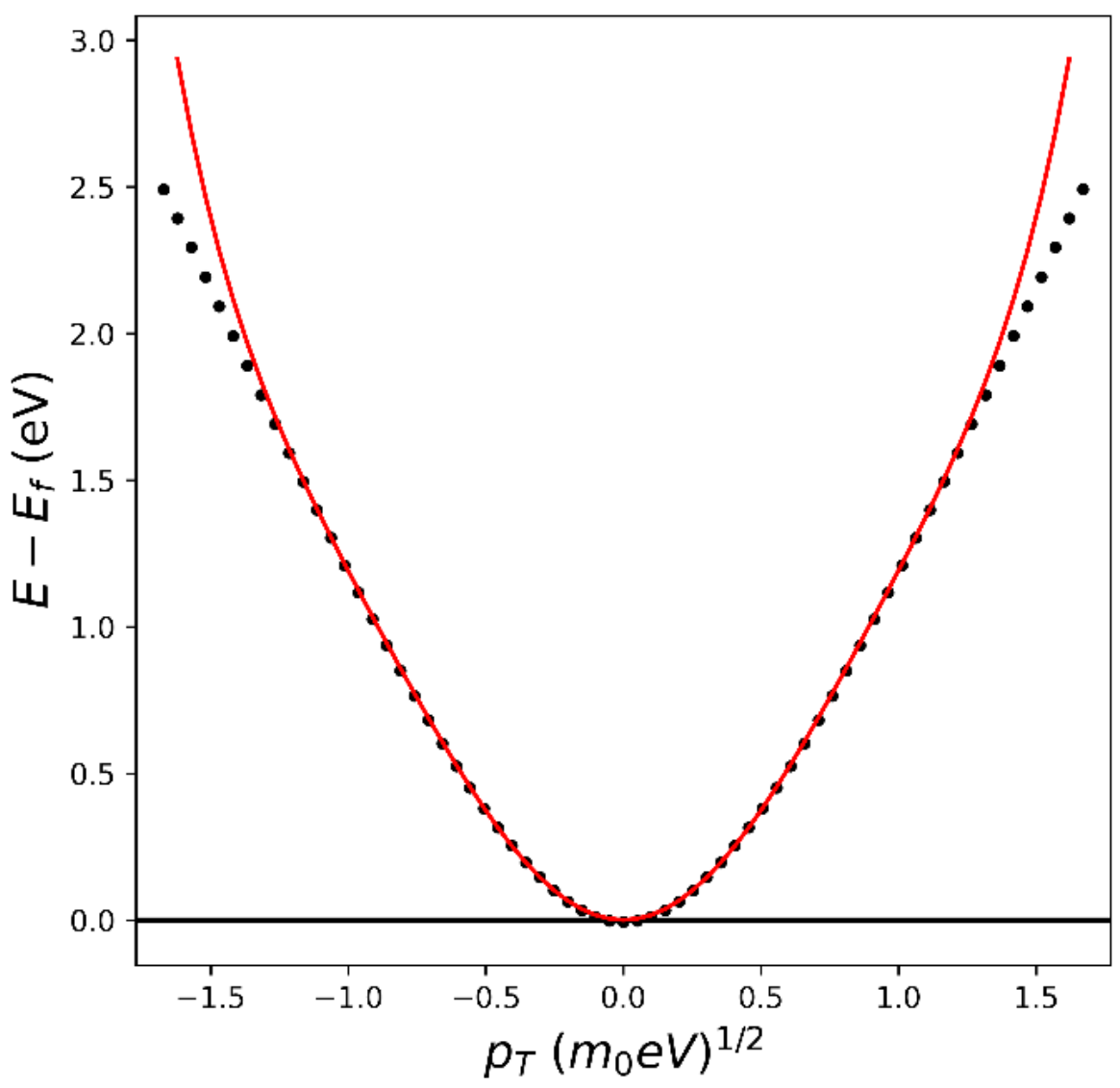

(a)

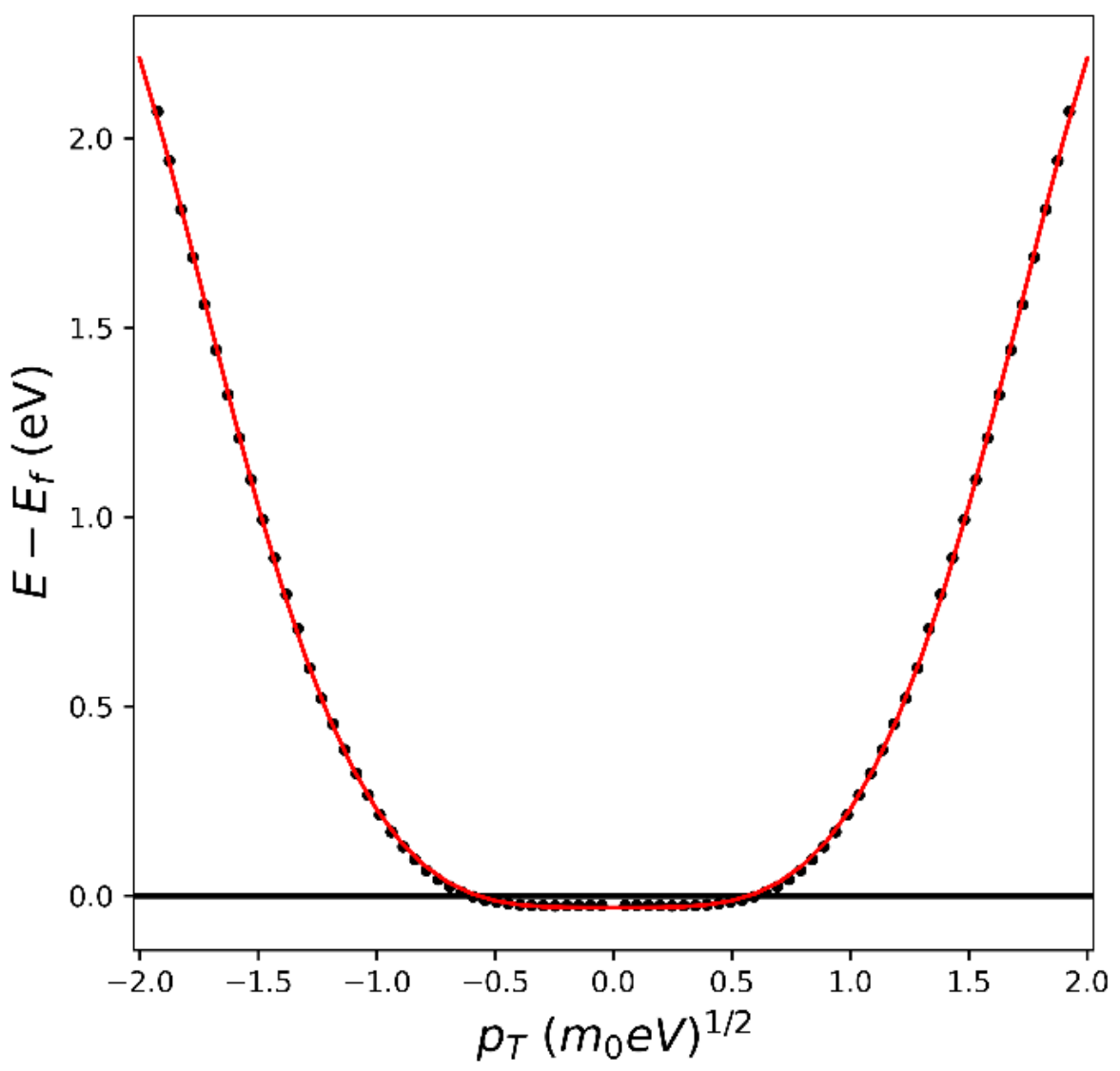

(b)

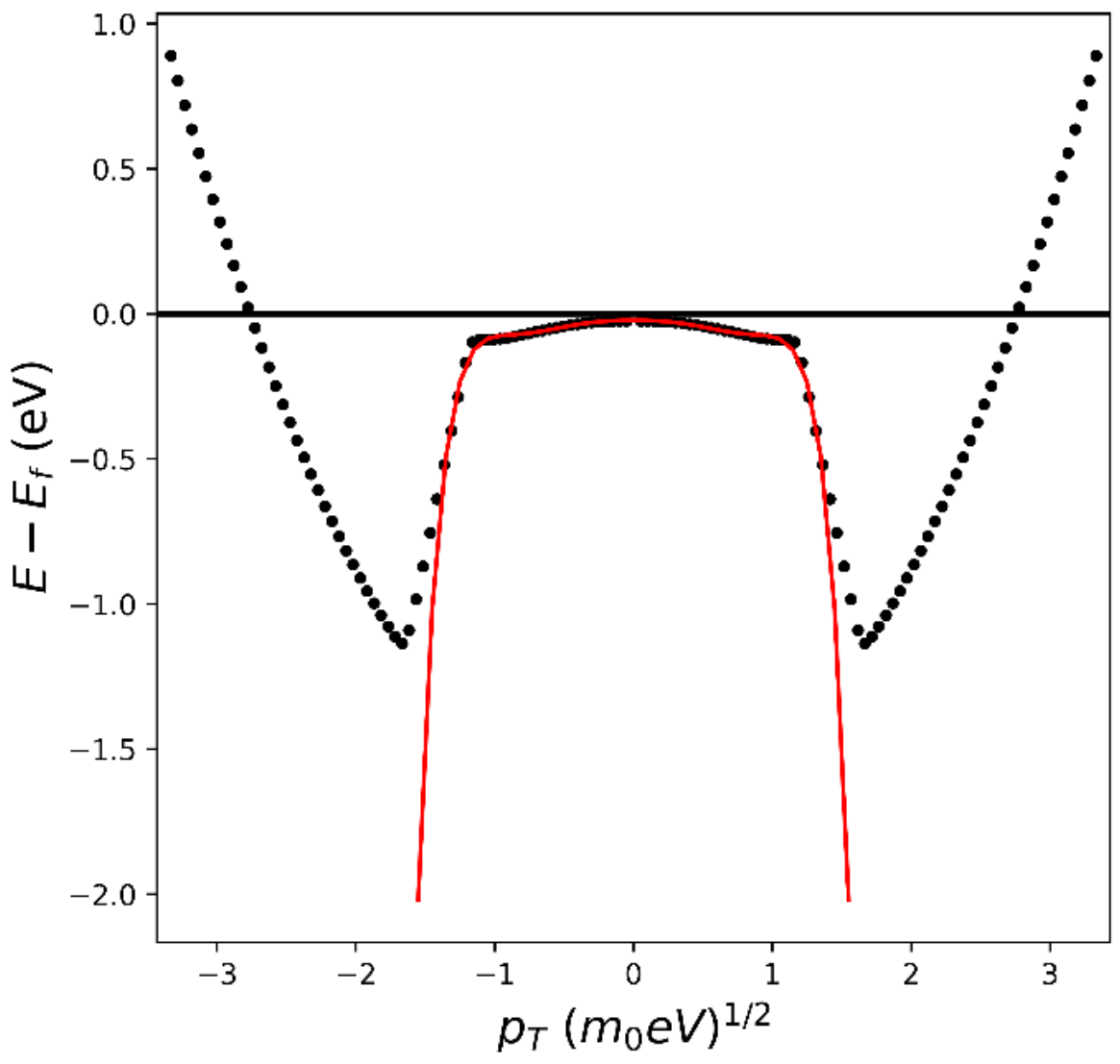

(c)

FIG 3. Polynomial transverse band fits (red lines) to the DFT evaluated dispersion (black dots) around the Fermi level; (a) for Cu(001), (b) the $(1\bar{1}0)$ $x$-direction for W(111), and (c) the $(11\bar{2})$ $y$-direction for W(111).

To simplify the numerical one-step photoemission simulation, the band dispersion in the longitudinal ($p_z$) emission direction is fit using a parabolic approximation for its energy-momentum relationship around the Fermi level; $E(p_z) = E_0 + \frac{(p_z - p_0)^2}{2m_z}$, where $E_0$ and $p_0$ are fitting parameters and $m_z$ is the effective mass of the band in the emission ($z$) direction. Figure 2 displays the longitudinal band fits (red lines) to the DFT-evaluated band dispersion (black points) for the single emitting band in both face-centered cubic Cu(001) and body-centered cubic W(111) photocathodes in the relevant Γ-X and Γ-P directions respectively. In the transverse direction around the Fermi level, a parabolic fit to the band dispersion proved insufficient, so a series fit of the form $E(p_T) = \sum_{k=0}^{n} c_k p_T^{2k}$ was used. These fits (red lines) to the DFT-calculated transverse band dispersions (black points) are shown in Figure 3. For Cu(001), the transverse dispersion of the emitting electron-like band is well-fit using $n = 3$ terms for excess photoemission energies $\Delta E$ less than about 1 eV (equivalent to transverse momenta $p_T < 1.4$ $(m_0.\mathrm{eV})^{1/2}$). For W(111), the transverse band dispersion is more complex, requiring separate fittings in the '$x$' $(1\bar{1}0)$ and '$y$' $(11\bar{2})$ directions, but the fits are still a good reflection of the transverse band dispersions for $\Delta E <$ 0.5 eV when terminating the series fit at $n_x = 3$ and $n_y = 4$. The numerical values of the longitudinal and transverse fit parameters $m_z$, $p_0$, $E_0$, and the $c_k$ coefficients for the emitting bands of the two single-crystal metal photocathodes are listed in Table I. We note that the employed band fittings are only accurate near the Fermi level as the $c_k$ series coefficients are strictly functions of $p_z$, and that the two-dimensional transverse dispersion of the emitting band in the Γ-P direction of W is more complex than described by a simple fit in two perpendicular directions [60]. Nonetheless, as photoexcited band states near the Fermi level and around $p_T = 0$ generally dominate photoemission for any excess energy $\Delta E$, due to their largest transmission probability over the work function barrier, the one-step photoemission simulation should provide a good approximation to direct band emission from below threshold to at least 0.5 eV above threshold – the spectral region of interest in this paper for both the Cu(001) and W(111) photocathodes.

TABLE I. Band fitting parameters for Cu(001) and W(111) photocathodes

| | Cu(001) | W(111) '$x$' | W(111) '$y$' |
|---|---|---|---|
| Longitudinal: | | | |
| $E_0$ [eV] | −9.3083 | −1.47444 | −1.47444 |
| $p_0$ [$(m_0.\mathrm{eV})^{1/2}$] | −2.006 | −0.213032 | −0.213032 |
| $m_z$ [$m_0$] | 1.998 | 0.65909 | 0.65909 |
| Transverse: | | | |
| $c_0$ [eV] | 0 | −0.0315134 | −0.0224986 |
| $c_1$ [$m_0^{-1}$] | 1.3141 | 0.000933322 | −0.0784615 |
| $c_2$ [$m_0^{-2}$ eV$^{-1}$] | −0.240957 | 0.298074 | −0.0125029 |
| $c_3$ [$m_0^{-3}$ eV$^{-2}$] | 0.0277 | −0.0395448 | 0.306276 |
| $c_4$ [$m_0^{-4}$ eV$^{-3}$] | – | – | 0.160303 |

For each crystal momentum point in the emitting band with energy $E(p_T,p_z)$, the Fermi-Dirac distribution multiplies the evaluated local DOS [12,60] to give the effective local electron population that emits with a longitudinal transmitted probability flux over the work function barrier given by

$$j_z = \frac{p_{z0}}{m_0}\left[\frac{4p_z p_{z0}}{(p_z + p_{z0})^2}\right], \tag{3}$$

where the longitudinal component of the emitted electron momentum $p_{z0}$ is evaluated using energy and transverse momentum conservation in photoemission. Summation over all emitting band $p_z$ values for each $p_T$ (i.e., $p_x$ and $p_y$) then gives the two-dimensional transverse momentum distribution of electrons emitted by one-step direct band photoemission at each $\Delta E$ [12,13]. Further summation over the evaluated electron emission distribution gives a measure of the QE as a function of $\Delta E$ which we normalize to unity at threshold ($\Delta E$ = 0) since the photoemission simulation does not include all physical effects (absorption depth, scattering, etc.) that affect the QE. Numerical calculation of the variance of the simulated two-dimensional $p_T$ distribution is, of course, directly related to the MTE. The numerical evaluations typically employ 70 equally spaced transverse momentum points in the $x$ and $y$ directions (from 0 to $p_{T,max} = \sqrt{2m_0(\Delta E + 20k_BT_e)}$) and 200 in the longitudinal direction (for $p_z$ values associated with energies from $\varepsilon_F - \Delta E$ to $\varepsilon_F + 20k_BT_e$) to ensure that the QE and MTE have a computational evaluation uncertainty of less than 1%. For values of $\Delta E$ less than −0.1 eV and highly dispersive electronic bands, the momentum limits of the numerical simulation are adjusted to assure the accuracy and convergence of both the QE and MTE evaluations.

## III. THEORETICAL UES EMSSION FRAMEWORK

The evident disagreement, especially near and below threshold for the W(111) photocathode, between the experimental data and current accepted photoemission models [7,8,12,13] for the spectral dependences of the QE and MTE displayed in Figure 1 strongly indicate the presence of an additional emission mechanism. Our proposed second contribution to photoemission from metal photocathodes builds on the work by Berglund and Spicer [14,65] in 1964 on Cu and Ag. These seminal papers provide the theoretical framework for the electron scattering analysis presented here with the exception that rather than analyzing elastic electron-electron collisions which only exchange momenta we invoke Umklapp electron scattering [49] (UES) of the form

$$\mathbf{p}_1^* + \mathbf{p}_2 = \mathbf{p}_0 + \mathbf{p} \pm \mathbf{G} \tag{4}$$

where a photoexcited electron of momentum $\mathbf{p}_1^*$ collides with an electron of momentum $\mathbf{p}_2$ (in an occupied bulk state) to generate, with the help of a reciprocal lattice vector $\mathbf{G}$, an emitted electron of momentum $\mathbf{p}_0$, and consequently a vacuum energy above the work function potential barrier of $E_{vac.} = p_0^2/2m_0$, leaving the second electron of momentum $\mathbf{p}$ in the bulk. In other words, the ‘inelastic’ nature of UES allows photoemission based on the momentum-resonant FC mechanism [44,48]. The 'strength' of UES in a metal may be estimated by determining the ratio of the metal's low temperature electron specific heat γ to that evaluated using the Sommerfeld free electron model for the metal $\gamma_0$. This ratio defines the thermal electron effective mass [66], $M_{th} = (\gamma/\gamma_0)m_0$, which is closely related to the optical electron effective mass in noble metals [67], and provides a measure of the degree to which inelastic electron-electron scattering perturbs electron heating in a metal.

Following the analysis of Berglund and Spicer [65], the initial 'Auger-like' collision process in UES involves a photoexcited electron (above the Fermi level) losing energy that is gained by an electron in the occupied bulk states. Before the collision and in the constant global DOS approximation, the initial photoexcited electron distribution is proportional to $f(E - \hbar\omega)$, where the Fermi-Dirac distribution $f(x) = (1 + e^x)^{-1}$, ħω is the photon energy, $x = (E - \hbar\omega)/k_B T_e$, and the Fermi level is defined as zero energy. After the collision, this electron upon losing an energy Δ must enter an unoccupied state – a probability proportional to $(1 - f(E - \hbar\omega + \Delta))$. Summation over all possible initial state energies gives a solvable integral of the form

$$\int_{-\infty}^{\infty} \frac{dx}{(1 + e^{-(x+d)})(1 + e^x)} = \frac{\ln[1 + e^d] - \ln[1 + e^{-d}]}{e^d - 1}, \tag{5}$$

where $d = \Delta/k_B T_e$. As this expression only describes the collisional de-excitation of the photoexcited electron, it must be convolved (in energy) with the distribution of the electrons residing in the occupied bulk states to determine the final distribution of the UES excited electrons. This convolution is performed numerically in our UES emission simulations, and we assume that the latter convolving Fermi-Dirac distribution is of the form $\left(1 + exp\left[m_0 N_{fs} E/(M_{th} k_B T_e)\right]\right)^{-1}$ to take into account empirically the inelastic nature of UES; that is, the additional inherent 'inertia' of UES due to the involvement of the reciprocal lattice vector **G** (equation (4)) in the inelastic scattering over the metal's Fermi surface(s) as 'quantified' by $M_{th}$. The number of Fermi surfaces $N_{fs}$ in the Brillouin zone is also introduced in this expression to account for the statistical difference between internal UES and the UES-mediated FC emission process described here; specifically, for the latter there is only one vacuum band to scatter into whereas for the former there are $N_{fs}$ Fermi surfaces to scatter into. For example, for two bulk Fermi surfaces ("1" and "2"), there are four possible internal Umklapp scattering events between the Fermi surfaces (1→1, 1→2, 2→1, and 2→2), whereas there are only two possible emission processes into the vacuum states (1→vacuum and 2→vacuum), implying a reduction in the strength of the UES emission process over internal UES scattering (as quantified by $M_{th}$) by a factor of $2 = N_{fs}$. The final distribution evaluated in this manner will then determine the number of electrons (and their energy) that are above the vacuum level following UES and so can be emitted.

The most probable emission process will be one that is both energy and momentum resonant; that is, a momentum-resonant Franck-Condon mechanism [44,48] in which the emitted electron satisfies the vacuum electron energy-momentum relation $E = p_0^2/2m_0$. Under these conditions, the components of the electron momentum perpendicular to the photocathode surface ($z$ direction) inside the bulk $p_z$ and outside in the vacuum $p_{z0}$ are equal, which allows the emitted electron flux $\mathbf{j}_T$ transmitted over the photoemission barrier to be written as [44]

$$j_T = \frac{p_{z0}}{m_0}\left[\frac{4p_z p_{z0}}{(p_z + p_{z0})^2}\right] = \frac{p_0 cos\theta}{m_0}\,, \tag{6}$$

where θ is the polar angle about the $z$ direction. Incorporation of the recipient vacuum DOS, proportional to $\sqrt{E}dE$ at energy $E$, further modifies the net transition probability of electrons into the vacuum.

We also propose that an additional resonant process is present for photoelectron emission via the UES-mediated FC process, one that is associated with the near zero vacuum group velocity of electrons emitted with very low excess photoemission energies. Such an effect has been observed in secondary electron emission spectroscopy of metals, for example in tungsten by R.F. Willis [68] and R.F. Willis and N.E. Christensen [69]. In their work, a resonance in emission was detected at energetic positions where the bulk band structure exhibits a minimum or maximum; that is, where $\frac{dE}{dp} = 0$, corresponding to quasi-stationary bulk states with zero group velocity. Willis et al. [69] also observed a similar effect at the emission threshold (i.e., around zero vacuum energy) which they attributed to a measurement artifact. We suggest that this threshold resonance is, at least in part, real as it can be attributed to the penetration of the near zero energy vacuum electron wave functions with $\frac{dE}{dp} \approx 0$ into the bulk of the solid-state photocathode. The likely energetic form of transition probability into the vacuum for this threshold resonance can be calculated by considering the time integral in standard time-dependent perturbation theory (for initial ($i$) and final ($f$) states); namely

$$\left|\int_0^\infty dt\, exp\left[i\left(E_i - E_f\right)t/\hbar - \eta t/2\right]\right|^2 = \frac{4\hbar^2}{4(E_{vac.})^2 + (\hbar\eta)^2}\,, \tag{7}$$

where the vacuum electron energy $E_{vac.} = p_0^2/2m_0$ is introduced, since the peak of the resonance is at zero vacuum energy and emission does not occur for $E_{vac.} < 0$, and $\eta$ is the inelastic electron scattering rate for the emitted electron at the vacuum level (i.e., at an energy equal to the value of the work function ϕ above the photocathode's Fermi level). For the considered case of FC electron emission, where the longitudinal electron momentum perpendicular to the photocathode emission face $p_z = p_{z0}$, the inelastic scattering rate for emission may be written as

$$\eta = \frac{p_{z0}}{m_0 \lambda}\,, \tag{8}$$

where λ is the inelastic mean free path at $E \approx \phi$ above the Fermi level in the metal. Here, we have used the $z$ component of the vacuum electron velocity as this reflects the ability of the electron to be emitted before its vacuum wavefunction penetrating into the photocathode is scattered in the bulk metal. In general, $\lambda(E \approx \phi)$ is only well known for a few materials, for example, for Cu where it is about 5nm [70,71], but reasonable values may be evaluated using the available literature for $\lambda(E \approx 0)$ at the Fermi energy [72] and the universal scaling law for inelastic electron scattering in materials [73]. Fortunately, due to the sharp Lorentzian resonance centered at $E_{vac.} = 0$ (equation (7)), a ±20% variation in the value of λ generates less than a ±5% change in the value of the MTE of electron emission from metal photocathodes for typical values of $\lambda(E \approx \phi) > 1$ nm; that is, significantly less than a similar change in the value of $M_{th}/N_{fs}$. As a gauge, equation (8) indicates that for the emission of electrons in the Boltzmann tail of a 300 K Fermi-Dirac distribution, $p_{z0} \approx \sqrt{2m_0k_BT_e}$, and an IMFP of 1 nm, the energetic width of the emission resonance is only around 60 meV

To evaluate the MTE of photoemission associated with UES, $MTE_{UES}$, we perform a numerical two-dimensional convolution of the spatial transverse electron momentum emission profiles simulated for direct band emission [12,13], $MTE_{direct}$, and FC-UES emission, $MTE_{FC}$. In other words, we make the ansatz that the two emission processes are not independent in the transverse direction as they both involve initial virtual excited states, albeit that direct band emission is only possible for the very small fraction of the total electronic excitation above the vacuum level in the reduced Brillouin zone. This convolution tends to generate near Gaussian total transverse momentum distributions with $MTE_{UES} \approx MTE_{direct} + MTE_{FC}$, especially near and below the photoemission threshold, in general agreement with experimental observations for metals with strong UES. On the other hand, the simulated QEs of the two emission processes are summed since the Hamiltonian for photoemission is discontinuous in the longitudinal emission direction (due to the potential step) but not in the transverse direction. Further, as our theoretical model does not include all the detailed physics of either the overall photoexcitation of electrons in the metal photocathode's band structure or that of the inelastic (and elastic) electron scattering processes, the photoemission simulation cannot determine the absolute QE of each emission process, only their dependence on the excess energy Δ$E$. In other words, the theoretical ratio of the direct band emission to that from the UES process is not known. Consequently, the simulated spectral dependence on Δ$E$ of the total QE, of the form $QE_{tot.} = A.QE_{direct} + B.QE_{UES}$ where $A$ and $B$ are constants and $QE_{UES} = QE_{FC}$, is fit to the experimentally measured QE(ħω) where ħω is the incident photon energy. To do so, the work function $\phi$ of the crystalline metal surface must be determined since $\Delta E = \hbar\omega - \phi$. As shown in Ref. [13] and by Gobeli and Allen [74], a plot of $(QE_{tot.})^{1/n}$ as a function of ħω yields a linear functional dependence for $n$ values generally between 2.5 and 3.5 and for $\Delta E > 0.1$eV at 300K to allow the work function to be extracted from the abscissa intercept; that is, using $(QE_{tot.})^{1/n} = C(\hbar\omega - \phi)$ where $C$ is a constant. This is in contrast to the MTE which is by its definition 'normalized' so that its theoretical spectral dependence (on Δ$E$) may then be directly compared to the experimental data using the extracted value of $\phi$ from the QE.

One important effect influencing the ratio of $QE_{direct}$ to $QE_{UES}$ can however be included; that is, the expected fraction of electrons emitted by one-step direct band emission to that by the FC-UES effect. This ratio is dependent upon the product of the UV radiation absorption coefficient α and the IMFP λ; specifically, it is straightforward to show that for one inelastic scattering event the fraction of photo-excited electrons emitted by one-step direct band emission is $\alpha\lambda/(1 + \alpha\lambda)$ and the fraction emitted by the FC-UES process is $1/(1 + \alpha\lambda)$, assuming that a to be emitted electron is headed towards the photocathode emission face upon which the UV radiation is incident

– our experimental case. As the optical properties of both Cu [67] and W [75] are well-known and a reasonable approximate value for $\lambda(E \approx \phi)$ can be evaluated for each metal [70-73], this relative emission efficiency ratio is included in our simulation of the QE. In most cases, the product $\alpha\lambda$ is relatively constant as the Lorentzian of equation (7) limits FC-UES emission to at and just above the vacuum level (fixing $\lambda(E \approx \phi)$) and, except for Cu [67], the UV absorption coefficient is generally relatively constant for metals around $\hbar\omega \approx \phi$. Of course, this QE ratio approximation may breakdown for metals in which $\alpha\lambda << 1$ due to multiple inelastic scattering events, but one can expect photoexcited virtual-state electrons scattered by more than one Umklapp scattering event to be lost to emission [7,14] as (i) most final state collision momenta will not be resonant with the vacuum states and (ii) their final energy state is more likely to be below the vacuum level.

## IV. COMPARISON OF THEORY TO EXPERIMENT

The above outlined FC-UES emission formalism is readily combined with our one-step direct band photoemission model [12] to allow comparison with the experimental measurements of the spectral dependence of both the QE and MTE. Table II lists the values of $M_{th}$, $N_{fs}$, and $\lambda$ used to simulate the FC-UES emission component for Cu and W. For W the UV absorption depth is relatively constant at 6 nm [75] over the 3.8-5.0 eV measurement range (Figure 1), whereas for Cu the variation in $\alpha$ with incident photon energy [67] was explicitly included. The fitted band dispersions employed to describe direct band photoemission for the Cu(001) and W(111) photocathodes are the same as those displayed in Table I.

TABLE II. Parameters for Umklapp scattering emission

| | $M_{th}$ | $N_{fs}$ | $\lambda$ (nm) |
|---|---|---|---|
| Cu | $1.36m_0$ | 1 | 5 |
| W | $3.5m_0$ | 1 | 1.96 |

The simulated total QE, including emission ratio associated with the absorption-IMFP product $\alpha\lambda$, is fit to the experimental data using the relationship

$$QE_{tot.} = \frac{C(\alpha\lambda\, QE_{direct} + f QE_{UES})}{1 + \alpha\lambda}\,, \qquad (9)$$

where $C$ is an overall fitting parameter and $f$ adjusts the strength of the UES emission contribution relative to direct one-step photoemission. For consistency and simplicity, in the simulation both $QE_{direct}$ and $QE_{UES}$ are set equal to unity at the photoemission threshold ($\Delta E = 0$) as their absolute emission efficiencies are not evaluated, so that the parameter $C(\alpha\lambda+f)/(1+\alpha\lambda)$ equals the measured QE at threshold. Once a fit to the total QE is established, the measured total MTE of the emitted electrons should be a good fit to

$$MTE_{tot.} = \frac{C(\alpha\lambda\, QE_{direct}\, MTE_{direct} + f\, QE_{UES}\, MTE_{UES})}{(1+\alpha\lambda)\, QE_{tot.}}\,, \tag{10}$$

where $MTE_{direct}$ is the one-step direct band emission contribution to the MTE [12] and, as noted above, $MTE_{UES}$ is from the numerical convolution of the simulated emitted electron transverse momentum distributions of the direct band and FC-UES emission processes as the two contributions (linked through the initial virtual excited state) are not considered independent in the transverse direction. In writing equation (10), the total MTE is assumed to be made up of two independent emission contributions ($i$ = 1,2), in keeping with the individual αλ absorption-IMFP ratios, and so may be written as the normalized quantity [50]

$$MTE_{tot.} = \frac{QE_1 MTE_1 + QE_2 MTE_2}{QE_1 + QE_2}\,. \tag{11}$$

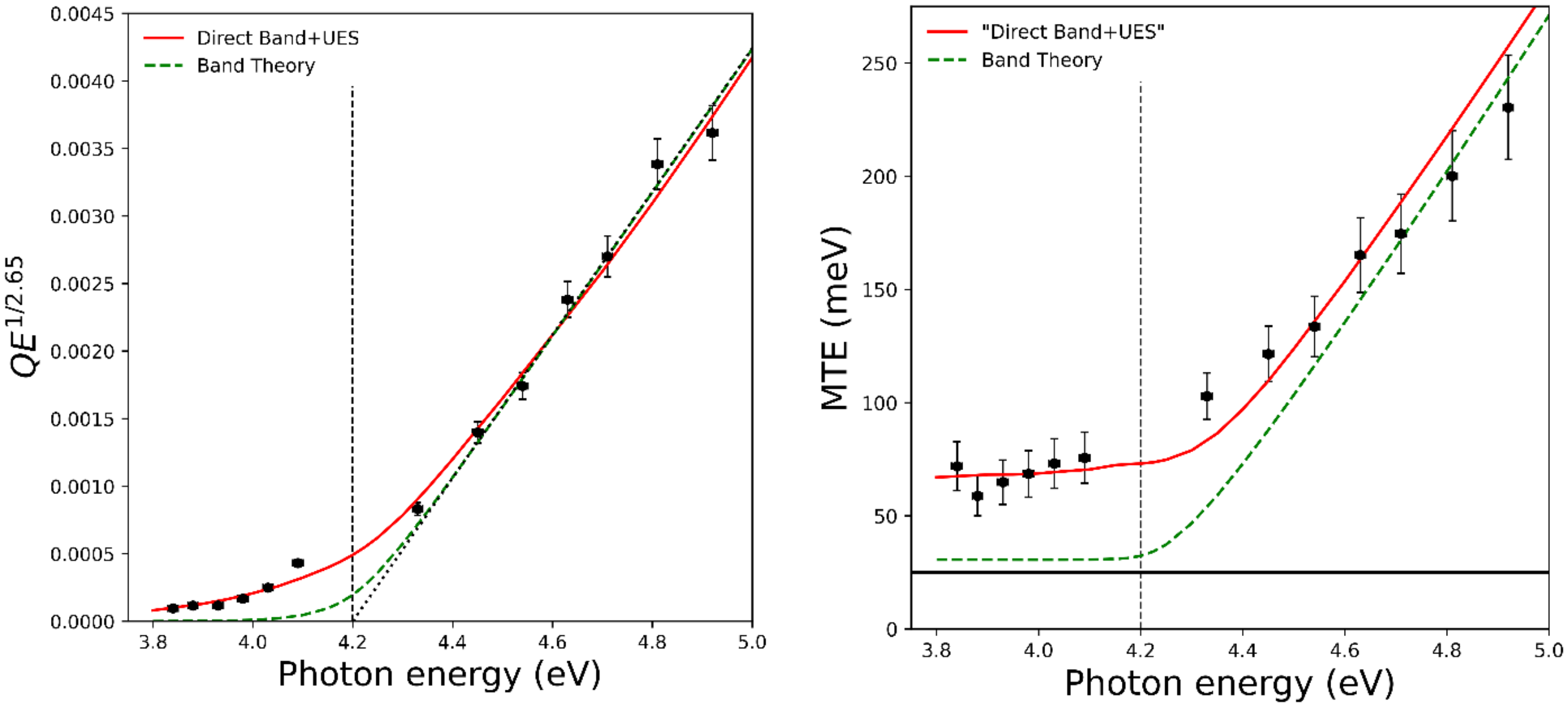

(a) (b)

FIG 4. Measured spectral dependences (black data points) for (a) the QE (plotted to the power of 1/2.65) and (b) the MTE as a function of the incident photon energy for the W(111) photocathode. The extracted work function of 4.2 eV from the $QE^{1/2.65}$ plot is shown by the dashed black line. The solid red lines display the results of the one-step direct band plus FC-UES photoemission simulation (see text). The dashed green lines show the one-step direct band emission and the horizontal black line at 25 meV in (b) gives the limiting below threshold MTE from the temperature-extended DS formalism for a 300 K photocathode [8].

Figures 4(a) and (b) compare the obtained theoretical fits (red lines) of the QE and MTE to the experimental data (black data points) for the W(111) photocathode. The QE is plotted to a power of $1/n$ (i.e., $QE^{1/n}$) against the incident photon energy ħω in Figure 4(a) for a value of $n$ = 2.65 which ensures a linear dependence for excess energies $\Delta E$ > 0.2 eV of the expected form $QE_{tot.}^{1/n} = C(\hbar\omega - \phi)$ [13,74]. As shown by the dotted line, the abscissa intercept implies a value of 4.2(±0.1) eV for the work function of the (111) crystal face of tungsten, which is at the lower end of the accepted literature value [76] but reasonably consistent with our DFT-based thin-slab calculations [77] that predict $\phi_{W(111)} = 4.0(\pm 0.2)$ eV. This value of $\phi$ is used to find $\Delta E$ for the plots in Figure 1 and to correlate the experimental measurements with the photoemission simulations which are evaluated as a function of $\Delta E$. We also note here that both the experimental data and theoretical simulations agree on the value of $n$ and that that value of $n$ is not 2; that is, the QE does not follow the $\Delta E^2$ dependence predicted by the temperature-extended DS formalism [7,8]. Again, this is primarily due to the $\sqrt{E}$ dependence of the vacuum DOS which is not included in the Vecchione-DS formalism.

The measured spectral dependence of $QE^{1/2.65}$ for the W(111) photocathode, plotted as the black data points in Figure 4(a), are quite accurately fit using $QE_{tot.}$ given by equation (9) (red line) with $f = 4$, the known relatively constant value of $\alpha^{-1} \approx 6$ nm over the investigated UV wavelengths [75], and the parameters listed in Table II. In particular, the direct band plus FC-UES emission simulation correctly describes the near and below threshold quantum efficiency. This is to be contrasted with the one-step direct band emission $QE_{direct}$ (dashed line in Figure 4(a)) which significantly underestimates the QE in this spectral region; that is, cannot explain photoemission physics. On the other hand, at high excess photoemission energies ($\Delta E > 0.3$ eV), both theoretical trends agree with the experimental data (including the $QE \propto \Delta E^{2.65}$ power law dependence), thus confirming that one-step direct band photoemission dominates the FC-UES emission process well above threshold.

As shown in Figure 4(b), our theoretical direct band plus FC-UES emission simulation of the MTE (red line) is also in good agreement with its measured spectral dependence for the W(111) photocathode (black data points). Importantly, the combined theory for $MTE_{tot.}$ embodied in equation (10) and using the same parameters as for the QE fit (Figure 4(a)) upholds the measured ~70 meV below threshold MTE, in stark contrast to the limiting value of ~30 meV predicted for $MTE_{direct}$ (dashed line). For $\Delta E$ greater than 0.3 eV, both $MTE_{tot.}$ and $MTE_{direct}$ are consistent with the experimental data, again indicating the dominance of one-step direct band photoemission. It is also notable that in this high excess energy regime that the MTE for the W(111) photocathode closely follows the $\frac{1}{3}\Delta E$ dependence of the Vecchione-DS formalism [7,8]. It is only near and below threshold that there is significant disagreement.

For the Cu(001) photocathode, the spectral $QE^{1/n}$ dependence is linear for $n = 3$ as shown in Figure 5(a), from which a work function of 4.1(±0.1) eV is recovered (dotted line) using the procedure of Gobeli and Allen [74]. This experimentally determined value of $\phi_{Cu(001)}$ is lower than both the 4.78(±0.2) eV value from our DFT-based thin-slab evaluation [77] and that of 4.59 eV derived from a survey of the literature [76]. Despite the employed aqueous chemical etching [54] and additional laser cleaning process [50,55-57], this disagreement may be due to the formation of stable hydroxyl layer that produces a theoretical work function reduction of ~0.5 eV [78] or possibly a (bulk)Cu-O-Cu surface dipole associated with a residual thin oxide layer on the Cu(001) surface. The experimentally extracted work function is of course used to determine $\Delta E$ for the plots in Figure 1 and to compare the spectral measurements of the Cu(001) photocathode with the photoemission simulations. Again, it is clear from Figure 5(a) that the QE does not follow the expected $\Delta E^2$ dependence of the temperature-extended Vecchione-DS photoemission formalism [7,8].

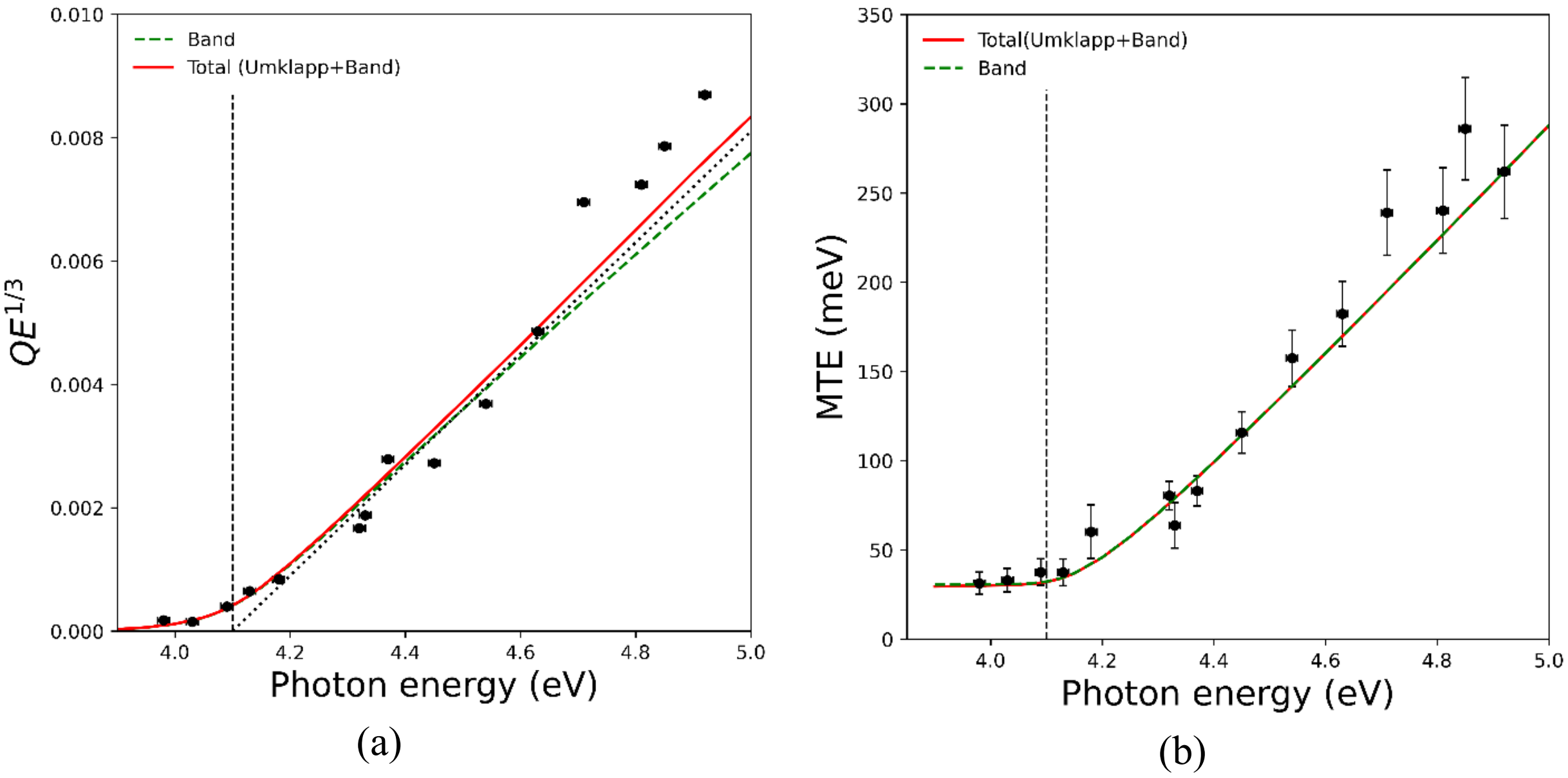

FIG 5. Measured spectral dependences (black data points) for (a) the QE (plotted to the power of 1/3) and (b) the MTE as a function of incident photon energy for the Cu(001) photocathode. The extracted work function of 4.1 eV from the $QE^{1/3}$ plot is shown by the dashed black line. The solid red lines display the results of the one-step direct band plus FC-UES photoemission simulation (see text) and the dashed green lines show the one-step direct band emission.

As is evident from Figure 5, for the Cu(001) photocathode, the theoretical photoemission simulations are in good agreement with the measured spectral dependences of both the QE and MTE (black data points). In this case, both the direct band plus FC-UES emission model (red line) and the one-step direct band emission simulation (dashed line) are consistent with the experimental data using the material parameters in Table II and the spectral dependence of the absorption coefficient [67]. Indeed, for the direct band plus FC-UES emission simulation, good agreement can be obtained for $f \leq 0.01$; the analysis for $f = 0.005$ is shown in Figure 5. This is in stark contrast to the results obtained for the W(111) photocathode near and below the photoemission threshold (Figure 4). The two main reasons for this are associated with the different material properties. First, the value of $\alpha\lambda$ is close to unity in Cu which means that emissions from the UES process are as likely as those from direct band emission: In W, $\alpha\lambda \approx 0.33$ and so the UES-based emission is 3-4× more likely by this metric. Secondly, the value of $M_{th}/(m_0 N_{fs})$ is also close to unity in Cu which means that the 'thermal tail' of the FC-UES emission mechanism is not substantially different from that of direct band emission: For W, $M_{th}/(m_0 N_{fs}) \approx 3.5$ so that the 'thermal tail' for UES-based emission is significantly longer, yielding a substantially greater $QE_{UES}$ (Figure 1(a)) and hence also the observed near and sub-threshold increase in the MTE (Figure 4(b)). For Cu(001), it is only well below threshold ($\Delta E < -0.2$ eV) that a divergence between the direct band plus FC-UES emission theory (red line) and the one-step direct band simulation of the MTE (dashed line) can be just observed for $f = 0.005$ in Figure 5(b). Nonetheless, both theoretical simulations are consistent with the experimental data that indicate a limiting below threshold MTE at 300 K for Cu(001) for about 29 meV, roughly 15-20% higher than that predicted by the temperature-extended Vecchione-DS formalism [7,8].

The main reason that the photoemission properties of the Cu(001) photocathode are in good agreement with the one-step direct band emission model is of course that, unlike tungsten, copper is much closer to a perfect 'Sommerfeld' metal with a near spherical Fermi surface [58,59] and $M_{th}$ close to the free electron mass [67] (Table II). Further, for emission from the (001) crystal face of Cu, the Γ-X direction in the Brillouin zone, there are no upper conduction bands near the vacuum level that could interfere with the observation of one-step direct band and FC-UES photoexcited electron emission [58,59], unlike for a Cu(111) photocathode [79]. Similarly, the band structure of tungsten [60] indicates that one-photon photoemission from a W(111) photocathode will also be unaffected by upper conduction bands. On the other hand, in addition to minor variations in modern DFT-based band structure calculations associated with choice of pseudo-potential etc., there is some uncertainty in the values of the material parameters ($M_{th}$, $N_{fs}$, and $\lambda(E \approx \phi)$) employed to describe the UES portion of the electron emission. We expect the resultant uncertainty in the simulated FC-UES mechanism to be less for Cu than for W as the literature values for these parameters are more certain for the former. For instance, the ±3% uncertainty in literature values for $M_{th}$ in copper [67,80,81] generates a 1-2% uncertainty in the sub-threshold MTE simulated using the direct band plus FC-UES emission analysis; that is, just greater than the ±1% uncertainty in the numerical simulation itself. In contrast, there is at least a ±15% uncertainty in $M_{th}$ for tungsten [60,82] which would translate to a ±10% uncertainty in the simulated sub-threshold value for the W(111) photocathode, giving about 70(±7) meV. Finally, we note that the values of $f$ employed to fit the experimental data are not related to the intrinsic strength of the UES process in the metal photocathodes – they are only fitting parameters associated with the ratio $QE_{UES}/QE_{direct}$ at $\Delta E = 0$ that is not due to the $\alpha\lambda$ product. Nonetheless, one should expect FC-UES emission efficiency enhancement due to $M_{th}$, but a more detailed investigation into the physical basis for this, and hence its functional form, is beyond the scope of this paper.

We also note that there is a declining agreement between theory and experiment for excess photoemission energies above 0.5 eV (i.e., photon energies above 4.6-4.7 eV); more so for Cu(001) (Figure 5) than for W(111) (Figure 4). This is, of course, primarily because the employed fits to the transverse dispersion of the emitting bands (Figure 3) are only accurate for $\Delta E < 0.5$ eV, with strong relativistic and spin orbit coupling effects in tungsten [60] likely causing larger deviations from the transverse band fits for occupied state energies further below the Fermi level. For copper, there is likely an additional effect. Band structure calculations [58,59] indicate an absorption onset at photon energies of 4.5-4.7 eV due to photoexcitation of electrons into an L-point upper conduction band with a minimum at around 4eV above the Fermi level [79]. As the extracted Cu(001) surface work function is 4.1(±0.1) eV (Figure 5(a)), rapid elastic momentum-exchanging electron-electron collisions between the photoexcited and bulk (ground state) electrons can readily produce electrons with momenta in the (001) crystal direction and at an energy above the vacuum level. The resultant additional emission process is consistent with the observed increase in the QE (Figure 5(a)), and some deviation from the theoretical spectral MTE variation (Figure 5(b)), for photon energies above 4.6-4.7 eV.

## V. DISCUSSION

The close agreement between the direct band plus FC-UES theoretical formalism and the spectral measurements for the Cu(001) and W(111) photocathodes for $\Delta E$ less than 0.5 eV is a strong indication that the additional photoemission mechanism required to understand the near and below threshold emission properties of metal photocathodes is related to Umklapp scattering [49]. The presented FC-UES emission formalism is, however, quite rudimentary in that it makes a number of simplifying approximations which can be improved upon.

First, the employed value of the thermal electron effective mass $M_{th}$ is obtained from a combination of literature values and calculations based on D. Gall's calculated electron mean free path [72]. For copper, this analysis gives consistent agreement with $M_{th} \approx 1.36m_0$. On the other hand, for tungsten, literature agreement with the calculated and employed value of $M_{th} = 3.5m_0$ is not as straightforward. Specifically, the ratio of the measured low temperature electron heat capacity to that evaluated using the free electron Sommerfeld model – the canonical definition for $M_{th}$ – gives a value of $2.6m_0$ assuming the two outermost 6*s* electrons/atom are free and using the corresponding value of 9.2 eV for the Fermi energy. On the other hand, our DFT band structure calculations for tungsten reveal that it is predominantly bands associated with the four 5*d* electrons that cross the Fermi level, suggesting that the Fermi energy is 14.6 eV and $M_{th} \approx 4.2m_0$. Both values of the Fermi energy are inconsistent with the value of 12.1 eV evaluated by Christensen and Feuerbacher [60]. However, Christensen and Feuerbacher also obtain a value of 0.86 mJ/(mol.K$^2$) for the electron specific heat using their relativistic/spin-orbit-coupling band structure calculations for tungsten, which for four free 5*d* electrons/atom then gives $M_{th} = 3.5m_0$ – the value we have employed in the presented FC-UES formalism. This suggests that accurate DFT-based band structure calculations could significantly enhance evaluations of $M_{th}$, providing both a value for the electron heat capacity for the actual Fermi surface and the number of free electrons per atom for Fermi energy evaluation. Moreover, as the DFT band-based electron heat capacity calculation does not include phonon effects, as it is a 0 K evaluation, the resulting value of $M_{th}$ obtained should be more reflective of just inelastic Umklapp scattering.

Second, the integer values for $N_{fs}$ used in the FC-UES simulation are estimated based on the number of major Fermi surfaces in the reduced Brillouin zone. Fundamentally, one would expect the Umklapp process contributing to photoemission to be dependent upon the occupied and unoccupied DOS involved in the inelastic scattering as restricted by its momentum conservation (equation (4)). Certainly, the total occupied DOS is directly involved in a DFT-based calculation of the electron specific heat $\gamma$ [60] and hence $M_{th}$ once the Fermi energy is known for an evaluation of $\gamma_0$ using the Sommerfeld free electron model [66]. However, the collisional momentum conservation for internal UES is different than that for the emissive FC-UES process since the emitted electron's vacuum momentum $\mathbf{p}_0$ is small compared to crystal momenta. Thus, for example, for a singular and purely spherical Fermi surface in a cubic metal crystal, equation (4) suggests that the Fermi momentum needs to exceed about $\frac{1}{3}\mathbf{G}$ to allow electron emission for a single scattering event. Such momentum conserving restrictions could eliminate certain Fermi surfaces or parts of Fermi surfaces from participating in FC-UES emission, implying that $N_{fs}$ may not be an integer if the described definition of $M_{th} = (\gamma/\gamma_0)m_0$ is to be used. Of course, contributions due to multiple Umklapp scattering events and those from electrons originating in neighboring Brillouin zones will also need to be considered.

Third, although not a major factor, the values of the IMFP at the vacuum level used in the FC-UES emission simulation could also be refined; that is, evaluated without the use of a universal scaling law [73]. In particular, Ridzel et al. [83] have shown that $\lambda(E \approx \phi)$ in metals can be directly related to their optical properties using the Mermin dielectric function. Their results agree with the use of 5 nm for copper but, in accord with H. T. Nguyen-Truong [84], suggest that $\lambda(E \approx \phi)$ is about a factor of 2 larger than the employed 1.96 nm for tungsten. In general, such differences only affect the ratio of direct band to FC-UES emission through the product $\alpha\lambda$, which we already take into account. However, we reiterate that only for $\lambda(E \approx \phi)$ less than 1 nm does the employed Lorentzian resonance at $E_{vac.} = 0$ becomes sufficiently broad to ensure that FC-UES emission more strongly influences both the total QE and MTE near and below the photoemission threshold. Niobium is a good example where this should be the case: In addition to its sub-nm IMFP at the vacuum level [85], it can be expected to have a thermal electron effective mass of $11.7m_0$ (calculated from the 5.23 eV Fermi energy and the 7.8 mJ/(mol.K$^2$) electron specific heat [82]) implying stronger FC-UES emission even with $N_{fs} \approx 2$ [86,87]. Further, for a Nb(001) photocathode, the primary emitting band in the Γ-H emission direction has a negative group velocity ($\frac{\partial E}{\partial p} < 0$) which may mean that the lifetime of its virtual excited state is shortened (due to faster break-up of the state for opposite group and phase velocities), potentially significantly reducing the QE of one-step direct band photoemission relative to FC-UES emission.

Finally, we note that the above-described simplified analysis of FC-UES photoemission allows for an approximate value for $MTE_{UES}$ to be written down for sub-threshold emission when $M_{th} \geq m_0 N_{fs}$ and $\alpha\lambda < 1$; that is, when photoemission through Umklapp scattering cannot be neglected. The convolution of the integral result of equation (5) with $\left(1 + exp\left[m_0 N_{fs} E/(M_{th} k_B T_e)\right]\right)^{-1}$ generates an exponential Boltzmann-like tail in the energy distribution of UES scattered electrons with a 'thermal' energy $k_B T \approx \left(M_{th}/(2m_0 N_{fs})\right) k_B T_e$. When combined with $MTE_{direct}$ through $MTE_{UES} \approx MTE_{direct} + MTE_{FC}$ (the approximate result of the MTE convolution), using $MTE_{direct} \approx \frac{6}{5} k_B T_e$ = 30 meV for $T = T_e$ = 300 K, one obtains

$$MTE_{UES} \approx \left(\frac{6}{5} + \frac{M_{th}}{2m_0 N_{fs}}\right) k_B T_e \,. \tag{12}$$

This expression is only valid for the case when there exists a single band crossing the Fermi surface with occupied states at transverse momenta $p_T$ up to about $\sqrt{40 m_0 k_B T_e}$; that is, the band dispersion does not restrict the available emitting band states [12,88]. For $M_{th} \leq m_0 N_{fs}$ and $\alpha\lambda < 1$, the opposite limit of weak FC-UES emission, direct band emission should dominate, and the below threshold MTE will then tend to ~30 meV at 300 K, as observed for the Cu(001) photocathode. For more than one emitting band or an emitting band that restricts the available transverse momentum states [12], the first term in the approximate expression for $MTE_{UES}$ (equation (12)) will need to be modified to reflect the restriction in $p_T$, differences in the dispersive properties of the emitting bands, and possible coherences between photoexcited virtual states (i.e., interference effects between photoemission transitions [89]).

## IV. SUMMARY

Experimental measurements on the spectral dependence of the QE and MTE of single-photon photoemission from single-crystal Cu(001) and W(111) photocathodes have provided evidence for an additional photoemission mechanism in metal photocathodes beyond that described by current theoretical analyses [7,8,12,13]. This additional mechanism is postulated to be based on an inelastic Umklapp electron scattering mediated Franck-Condon process and its presence is predominantly observed near and below the photoemission threshold, as direct one-step band emission dominates well above threshold. A simplified analysis of the proposed FC-UES emission process, involving only the thermal electron mass ($M_{th}$), the inelastic mean free path (IMFP) of electrons near the vacuum level ($\lambda(E \approx \phi)$), and the number of Fermi surfaces in the metal ($N_{fs}$), when included in a direct one-step band-based photoemission simulation [12] is consistent with the experimental QE and MTE data. This agreement is obtained with minimal additional fitting parameters once the ratio of direct to FC-UES emission (dependent on the product $\alpha\lambda$) is included; specifically, only adjustment of the relative strengths of the two emission processes is required, as the absolute emission efficiency of each is not simulated.

Definitive verification that the additional emission process is indeed a FC-UES process will require more detailed theoretical analysis and supplementary directed experimental investigations. Further theoretical studies will need to establish the strength of the FC-UES emission process (with respect to other photoemission processes) through a full simulation of the inelastic Umklapp electron scattering mechanism over the bulk band states and Fermi surfaces to provide reliable values for $M_{th}$ and $N_{fs}$. The physics of the resonant emission effect [69] around $E_{vac.} = 0$ will also need to be analyzed in more detail; specifically, around the value of the electron IMFP ($\lambda$) at this energy and whether the group velocity of the scattered emitting electronic state is also involved.

On the experimental side, spectral investigations of a single-crystal metal photocathode that does not allow one-photon direct one-step photoemission (or upper conduction band photoexcited thermalized emission [79]) for photon energies around the surface work function (i.e., $\hbar\omega \approx \phi$) could allow the study of the FC-UES emission process in isolation. Due to its unique electronic band structure [90-92], a Be(0001) photocathode is potentially ideal for such an investigation as it has no emitting bulk band states crossing the Fermi level between the Γ and A points of the Brillouin zone (the *c*-axis emission direction). Further, the nearest occupied band state in the perpendicular Γ-K/M plane of the Brillouin zone is at a transverse energy $p_T$ greater than 2 $(m_0.eV)^{1/2}$ from the Γ point, implying that bulk electron emission will only be possible via the FC-UES mechanism for photon energies below the expected 4.98 eV surface work function [93] to more than 7 eV – a spectral window that includes the ~6 eV fourth harmonic of Ti:sapphire lasers. Interestingly, although the QE is expected to be quite low in this spectral region, our theoretical simulations (using $\lambda(E \approx \phi) \approx 8$ nm parallel to the *c*-axis [72], $N_{fs} = 2$ [89], and a thermal electron effective mass [94] $M_{th} \approx 0.3m_0$ (also calculated [67] from the measured $\gamma = 0.171$ mJ/(mol.K$^2$) [82] and known 14.3 eV Fermi energy of Be)) indicate that the MTE of FC-UES emission (i.e., $MTE_{FC}$) may be well below 5 meV for $\hbar\omega \approx \phi$ at a 300 K lattice temperature; that is, significantly sub-thermal.

## Acknowledgements

This work was supported by the U.S. Department of Energy under Award no. DE-SC0020387.

## Data Availability

The experimental data supporting the findings of this study are openly available at https://indigo.uic.edu/articles/dataset/Tungsten_and_Copper_Photocathode_data/31990251/1 .

## Author Contributions

All three authors contributed equally to this work.